\def\year{2021}\relax
\documentclass[letterpaper]{article} 
\usepackage{aaai21}  
\usepackage{times}  
\usepackage{helvet} 
\usepackage{courier}  
\usepackage[hyphens]{url}  
\usepackage{graphicx} 
\usepackage{natbib}  
\usepackage{caption} 
\usepackage[draft]{hyperref}
\usepackage{graphicx,balance,multirow,xspace,subcaption,longtable}
\usepackage{makecell,amsmath,cleveref,rotating,amssymb}
\usepackage[font={small}, textfont=md]{caption}
\usepackage[T1]{fontenc}
\usepackage[draft,inline,nomargin,index]{fixme}
\usepackage{booktabs}
\usepackage{xcolor}
\usepackage[utf8]{inputenc}
\usepackage{marginnote}

\usepackage[colorinlistoftodos]{todonotes}

\PassOptionsToPackage{hyphens}{url}
\newcolumntype{P}[1]{>{\arraybackslash}p{#1}}
\newcolumntype{X}[1]{>{\centering\arraybackslash}p{#1}}

\expandafter\def\expandafter\UrlBreaks\expandafter{\UrlBreaks
  \do\a\do\b\do\c\do\d\do\e\do\f\do\g\do\h\do\i\do\j%
  \do\k\do\l\do\m\do\n\do\o\do\p\do\q\do\r\do\s\do\t%
  \do\u\do\v\do\w\do\x\do\y\do\z\do\A\do\B\do\C\do\D%
  \do\E\do\F\do\G\do\H\do\I\do\J\do\K\do\L\do\M\do\N%
  \do\O\do\P\do\Q\do\R\do\S\do\T\do\U\do\V\do\W\do\X%
  \do\Y\do\Z}

\crefformat{section}{\S#2#1#3}
\crefformat{subsection}{\S#2#1#3}
\crefformat{subsubsection}{\S#2#1#3}

\newcommand\clearrow{\global\let\rowmac\relax}

\newcommand{\para}[1]{{\vspace{.05in} \bf \noindent #1 }}
\newcommand{\parait}[1]{{\vspace{.05in} \em \noindent #1 }}

\newcommand{\subreddit}[1]{\emph{r/#1}}

\newcommand{\D}[1]{$\mathfrak{D}_{#1}$}

\renewcommand{\footnotesize}{\scriptsize}

\newcommand{\eg}{e.g.,\ }
\newcommand{\etal}{et al.\xspace}
\newcommand{\ie}{i.e.,\ }

\title{The Morbid Realities of Social Media: An Investigation into the
Misinformation Shared by the Deceased Victims of COVID-19}

\author {
  Hussam Habib,
  Rishab Nithyanand \\
}
\affiliations {
  University of Iowa \\
  hussam-habib@uiowa.edu, rishab-nithyanand@uiowa.edu
}

\begin{document}

\maketitle

\sloppy

\begin{abstract}
Social media platforms have had considerable impact on the real world especially
during the Covid-19 pandemic. Misinformation related to Covid-19 might have
caused significant impact on the population specifically due to its association
with dangerous beliefs such as anti-vaccination and Covid denial. In this work,
we study a unique dataset of Facebook posts by users who shared and believed in
Covid-19 misinformation before succumbing to Covid-19 often resulting in
death. We aim to characterize the dominant themes and sources present in the
victim's posts along with identifying the role of the platform in handling
deadly narratives. Our analysis reveals the overwhelming politicization of
Covid-19 through the prevalence of anti-government themes propagated by
right-wing political and media ecosystem. Furthermore, we highlight the failures
of Facebook's implementation and completeness of soft moderation actions
intended to warn users of misinformation. Results from this study bring insights
into the responsibility of political elites in shaping public discourse and the
platform's role in dampening the reach of harmful misinformation.
\end{abstract} 

\section{Introduction}
The influence of social media in shaping our perception of sociopolitical
realities is undeniable. It has long been understood and celebrated that their
algorithms facilitate the democratization of content --- enabling the
amplification of otherwise unheard voices. 
In recent times, however, this has led to the amplification of problematic and
harmful {misinformation} \cite{amarasingam2020qanon,fisher2016pizzagate}.
Indeed, recent findings show that their algorithms have been the target of
effective, intentional, and inorganic manipulation efforts to promote a variety
of conspiracy theories, erode trust in institutions, and generally sow
political disharmony \cite{Kremlin-Wapo22}. 
As will be demonstrated in this paper, the consequences of such manipulation,
unfortunately, transcend the virtual world and impact real people.

The abundance of misinformation was once again observed in the context of the
Covid-19 pandemic where anti-establishment, anti-vaccination, and anti-science
conspiracy theories were rampant \cite{united2020tackles}.
In fact, in 2020, the World Health Organization declared the abundance of
information surrounding the Covid-19 pandemic, including misinformation and
news from untrustworthy sources observed in digital media, as an
\emph{infodemic}. An infodemic is defined as false or misleading information
that can cause confusion and risk-taking behaviors with adverse effects to
health \cite{Infodemic}.
Although the exact harms of this infodemic are not measurable, it is noteworthy
that: (1) the United States alone lost more than 1.05M individuals to the
pandemic of whom nearly 100K were unvaccinated \cite{johnson2022covid}
\footnote{For clarity: This includes individuals who succumbed to Covid-19
prior to the vaccine. According to the CDC, unvaccinated individuals are
upto 53$\times$ more likely to succumb to Covid-19 \cite{johnson2022covid}.};
and (2) there have been many anecdotal reports linking these (likely
preventable) deaths to the digital infodemic \cite{wellness-wapo,
npr-momdied}.

Prior work has largely sought to characterize the types and prevalence of
various types of Covid-19 misinformation on different platforms or to
understand the effectiveness of platform moderation strategies at mitigating
their harms. 
What is lacking, however, is a systematic effort to characterize the
real-world impacts of different types of misinformation and the entities that
were responsible for amplifying their real-world harms. Our work seeks to fill
this gap.
Addressing this gap in research is important for several reasons: 
First and most broadly, it provides a deeper understanding of online trust,
influence, and manipulation. Second, it provides insights into the specific
epistemes that underlie misinformation that is capable of manipulation of
beliefs or proves useful in the rationalization of already conceived beliefs.
Finally, such a characterization might help platforms, influencers, regulators,
and citizens better mitigate the harms from future infodemics.

\para{Our contributions.}
\emph{In this paper, we aim to characterize the misinformation that was
present on the social media profiles of the deceased victims of Covid-19.} 
In other words, we analyze the content shared by a large sample of the
population that shared Covid-19 misinformation and then succumbed to Covid-19.
The goals of this analysis is threefold: (1) to shed light into
the misinformation themes that were espoused and amplified by the eventual
victims of the infodemic, (2) to identify the sources (\emph{elites} and
domains) responsible for creating and propagating the associated misinformation
themes, and (3) to identify the role of platforms in amplifying these
deadly narratives. We expand on each of these below.

\parait{Covid-19 misinformation themes (\Cref{sec:narratives}).} 
We use a custom word-embedding to cluster textually and semantically similar
posts. Next, we use random samples of posts from each cluster to manually label
the misinformation theme associated with the cluster. 
In our analysis, we focus specifically on measuring the prevalence of various
themes and characterizing how Covid-19 victims navigated between these themes. 
Measuring the prevalence of each theme allows us to better understand the
epistemes in the infodemic that were were used to: (1) manipulate victims'
beliefs about the pandemic or (2) help rationalize their pre-existing beliefs
about the pandemic.
Analyzing the patterns of victims' navigation between themes helps us
understand how a user might have adopted or propagated their harmful beliefs.
Our results show strong evidence of the harms of the politicization of the
pandemic with 67\% of all victims sharing anti-government themed misinformation
(\textit{vs.} 25\% anti-vaccination, 27\% anti-science).

\parait{Sources of Covid-19 misinformation (\Cref{sec:sources}).}
Victims often share posts from prominent public figures or link to content in
domains outside the platform. When this occurs, we refer to the public figure
or the external domain as the `source'.
In our analysis, we focus on characterizing the sources of misinformation
observed in our dataset. We use public data sources to: (1) group public
figures by their occupation and political stance and (2) group external domains
by their bias and credibility. We then measure their prevalence and prominence
in our dataset, and identify the misinformation themes they propagate.
This characterization allows us to identify the entities responsible for
amplifying specific misinformation themes. 
Our results once again emphasize the harms of politicization of the pandemic.
Specifically, we see that the right-wing political and media ecosystem were the
largest amplifiers of misinformation shared by the Covid-19 victims in our
dataset.

\parait{The role of platforms (\Cref{sec:platforms}).}
Facebook, along with other social media platforms, announced the application of
soft moderation (\eg warning labels) on Covid-19 related posts and
misinformation. 
In our analysis, we measured the completeness and consistency of platforms'
application of `soft moderation' labels on victims' misinformation posts.
We then breakdown this analysis by sources and misinformation theme to identify
specific gaps in the application of these moderation interventions.
Our results show that Facebook only applied soft moderation on 6\% of all
misinformation posts. However, we find that the application of these
interventions is consistent --- \ie similar posts are equally likely to obtain
the same intervention. We find that verified users have a marginally higher
likelihood of being the target of an intervention than non-verified users.

Taken all together, our analysis points to two key associations with the harms
caused by the infodemic: the politicization of Covid-19 by the political elite
and platforms' failure to effectively enforce interventions on Covid-19
misinformation.

\para{Caveat.}
It is important to note that we cannot and do not conclude that specific
misinformation themes (or entities) were causally responsible for the
real-world harm inflicted on the victims. 
Rather, we claim that these misinformation themes (or entities) were either
(1) causally responsible for the victims' harmful beliefs, or (2) used by
victims to support or rationalize their already harmful beliefs.
Put another way, our observational study can only yield correlated
relationships (not causal relationships).

\section{Data Collection}\label{sec:dataset}
We perform our study on crowd-sourced collections of case studies of misinformed
victims of Covid-19 posted online. These case studies contain posts of Facebook
users, who openly declared their anti-vaccination, anti-mask, and anti-science
beliefs online before succumbing to Covid-19 themselves. These posts present an
opportunity to closely study the topics, sources, and the reactions of the
platform that were associated with the beliefs that caused real-world harm. We
source these collections from two sources: \subreddit{HermanCainAward} and the
website \url{www.sorryantivaxxer.com}. Both of these sources were designed for
users to share stories about people who have made public declaration of their
anti-vaccination, anti-mask or Covid-hoax beliefs followed by contracting
Covid-19, such as the political figure Herman Cain, the namesake of the
subreddit \subreddit{HermanCainAward}. Therefore, they offer a unique and rare
insight into the Covid-19 misinformation associated with real-world victims of
Covid-19.
It should be noted that these communities have been the subject of significant
controversy due to their initial focus on schadenfreude. However, following
public criticism and moderator actions, they have aligned their goals towards
education of the harms of Covid-19 misinformation. We discuss our ethical
considerations related to the use of these datasets in \Cref{sec:discussion}. 

\para{Datasets.}
We used the Reddit PRAW API \footnote{\url{https://praw.readthedocs.io}} to gather all
submissions made to \subreddit{HermanCainAward} between 08/21 and 02/22. Each
submission contains a collection of images associated with a single victim of
Covid-19 and each image is a curated screenshot showing evidence of the
victim's belief in Covid-19 misinformation. In total, we gathered 1.7K unique
submissions containing 17.5K curated screenshots from Reddit.
We used a cURL-based crawler to scrape all `stories' published on the 
\url{sorryantivaxxer.com} website. Each story contains a time-ordered
collection of Facebook screenshots showing evidence of the victim's belief in
Covid-19 misinformation. From \url{sorryantivaxxer.com}, we gathered 281 unique
stories and extracted 3.2K curated screenshots associated with them.
In total, our collection included 1.97K unique victims and 20.8K screenshots
associated with their Covid-19 misinformation-related beliefs. We then used an
Optical Character Recognition (OCR) tool (EasyOCR) to recognize and extract all
text contained within the image. We conducted our analysis on this dataset of
users, screenshots, and texts.

\section{Narratives in COVID-19 Misinformation}
\label{sec:narratives}

\textbf{Overview.} 
In this section, we focus on: \Cref{sec:narratives:themes} identifying the
themes of Covid-19 misinformation seen in our dataset and uncovering how
victims progressed through these themes; and \Cref{sec:narratives:entities} the
entities and sentiments referenced in Covid-19 misinformation.

\subsection{Covid-19 misinformation themes} \label{sec:narratives:themes}
We seek to answer the question: {\em What are the themes associated with
Covid-19 misinformation posted by victims and how do they chronologically
progress through these themes?} At a high-level, we use clustering based on
semantic similarity in conjunction with manual cluster labeling to identify
misinformation themes and harness the chronological ordering of screenshots
associated with each victim to identify common theme progression patterns.

\para{Methods.} We now detail our methodologies for identifying themes and
uncovering common progression patterns.

\parait{Identifying misinformation themes.} We assign themes to misinformation
using the following three steps.
\begin{itemize}
  \item {\em Building a custom word embedding.} We begin by constructing
    a custom word embedding over the text extracted from all 20.8K screenshots
    in our dataset. We create this custom embedding using FastText
    \cite{fasttext} for two key reasons: First, the vocabulary associated with
    Covid-19 misinformation is unique and not captured in off-the-shelf
    embeddings. Second, we wish to preserve semantic {\em and} textural
    similarity between words. Maintaining semantic similarity means that words
    with lower distance between them in the word embedding are more
    semantically similar than those further away from each other. Maintaining
    textual similarity means that words with lower distance between them are
    more syntactically similar than words further away from each other ---
    simply put, {\em similarly} (mis)spelled words are closer together than
    those that are not. This latter requirement is especially important in our
    process because our reliance on OCR to extract text from screenshots
    may result in the creation of text spelling errors. We use FastText
    specifically because it achieves both requirements. 

  \item {\em Keyword-based post clustering.} Next, we generate TF-IDF vectors
    from the text associated with each post (using all 20.8K posts as
    the document corpus). These vectors provide a measure of importance (\ie
    {\em weight}) to each word in the post. We use these weights to compute the
    weighted mean coordinates of the post in our custom embedding. 
    For example, consider a post $S$, let $E(w)$ represent the coordinates
    associated with the word `$w$' in the custom embedding, and $T(w)$
    represent the TF-IDF weight associated with the word $w$ in $S$. Then, we
    compute $\frac{\sum_{w \in S}E(\text{w}) \times T(\text{w})}{|S|}$ as the
    weighted mean coordinates of $S$. 
    This weighting allows us to place more emphasis on the words determined to
    be more important to a given post.
    We then use simple k-means clustering to cluster the weighted mean
    coordinates of all 20.8K posts in our dataset. We use cluster coherence
    metrics and the elbow method in conjunction with manual validation to
    settle on $k=44$ for our dataset.

  \item {\em Manual label assignment.} Finally, we randomly sample 12 posts
    from each cluster and use an expert to determine the theme of the
    misinformation contained in each cluster. We then apply this label (\ie
    theme) to all posts within the cluster.

\end{itemize}

\parait{Measuring theme progression.} In order to understand the
relationships between themes in our dataset, we create convert our data into
a directed graph representation.
In this directed graph, we represent each theme as a node and use directed
edges to denote the chronological ordering of themes observed for each user in
our dataset. Thus, the weight on the directed edge from one node ($n_1$) to
another ($n_2$) represents the number of users who posted misinformation of
theme $n_1$ and immediately followed it by theme $n_2$. 
Next, to group topics that are more connected with each other, we perform the
Louvain method for community detection \cite{louvain}. This identifies cliques
with densely connected nodes. In the context of our analysis, these cliques
represent misinformation themes at are contiguous and compatible with each
other.

\para{Results.} In total, we identified 14 unique themes of misinformation
ranging from anti-government to alternative medication. Disturbingly, 82\% of
all users made posts regarding their own death and over 30\% made statements of
regret or pleas for help. 
A full list of the themes and their prevalence across victims and posts is
shown in \Cref{tab:themes}.
Our analysis shows that political (anti-government and anti-democrat) themes
which suggested a lack of trust in the (state or federal) government or the
democratic party occurred for significantly more victims (and posts) than
anti-science, anti-vaccination, conspiracy, or alternative medication themes.
From our chronological analysis, we find that over 38\% of all posts were made
after contracting Covid-19.
\begin{table}[]
  \centering
  \resizebox{0.6\linewidth}{!}{%
    \begin{tabular}{@{}llll@{}}
    \toprule
    \textbf{Topic}  & \textbf{\% of victims} & \textbf{\# posts} \\ 
    \midrule
    \textit{Anti-Government}       & 67          & 3040     \\
    \textit{Anti-Science}          & 27          & 750      \\
    \textit{Anti-Vaccination}      & 25          & 727      \\
    \textit{Anti-Democrat}         & 21          & 544      \\
    \textit{Masks}                 & 21          & 564      \\
    \textit{Religion}              & 19          & 532      \\ 
    \textit{Alternative Solutions} & 17          & 439      \\
    \textit{Mandates}              & 15          & 371      \\
    \textit{Immigration}           & 9           & 209      \\
    \textit{General COVID}         & 8           & 170      \\
    \textit{Pro-Freedom}           & 6           & 134      \\
    \textit{Conspiracy}            & 5           & 105      \\
    \textit{Quarantine}            & 1           & 17       \\
    \midrule
    \textit{Death}                 & 82          & 2725     \\
    \textit{Regret}                & 31          & 1174     \\
    \textit{Asking for Help}       & 30          & 883      \\
    \bottomrule
    \end{tabular}}
    \caption{Prevalence of themes in posts and victims. The bottom three rows
    denote non-misinformation themes and their prevalence.}
    \label{tab:themes}
    \end{table}

\begin{figure}
    \includegraphics[width=\columnwidth]{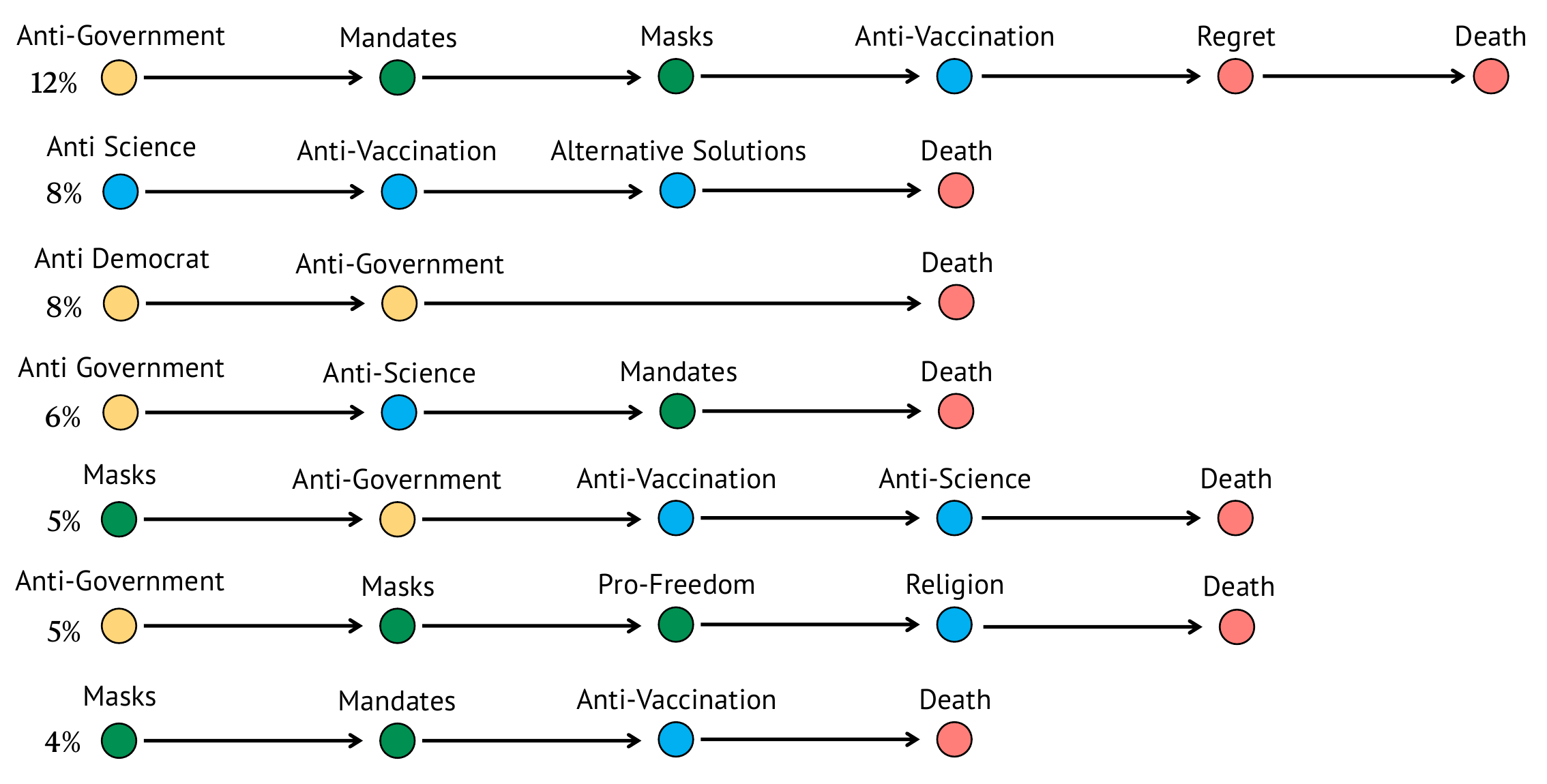}
    \caption{Representation of the most common progressions through narratives
    along with the percentage of users with similar progression. Colors of the
    nodes represent the cluster each topic belongs to based on the Louvian
    community analysis.}
    \label{fig:progression}
\end{figure}

From our theme progression analysis, we observe several common patterns in user
engagement of Covid-19 misinformation. First, we find that anti-masks/mandates
and pro-freedom themes were most commonly observed as a gateway to other
themes. This is likely capturing the victims' initial reactions to Covid-19
lockdowns and mandates. Conversely, the most frequently observed terminal
misinformation stages included anti-government and anti-vaccination posts. 
The themes found to co-occur least with other themes were religion and
alternative solutions/medications.
Second, nearly 63\% of all victims in our dataset appeared to follow a similar
progression through misinformation themes. These are illustrated in
\Cref{fig:progression}. We find the most common cohort to include multiple
anti-government themed posts followed by posts on masks and mandates.
Alternatively, we also observe users sharing a sequence of posts initiating
with anti-Democrat posts and then generalizing to anti-government topics. In
some cases, users tend to post anti-government posts followed by
anti-vaccination posts as well suggesting some form of influence between
anti-government and anti-vaccination beliefs. 
Finally, the biggest cohort without any anti-government themes included
intellectual narratives starting from anti-science posts to anti-vaccination
and finally alternative solutions. Users posting alternative solutions were
also found to be the least likely to share COVID-19 posts with regret.

\subsection{Covid-19 misinformation entities} \label{sec:narratives:entities}
In this section, we seek to answer the question {\em What are the entities
associated with Covid-19 misinformation posted by victims, and what are the
sentiments towards these entities?} Using
part-of-speech tagging and sentiment analysis, we seek to identify the prominent
entities present in our dataset, measure their distribution in the identified
Covid-19 misinformation themes and measure the sentiments associated with them.

\para{Methods.} We now detail our methodologies for identifying entities and
measuring their associated sentiment.

\parait{Identifying entities.} We identify common entities present in our
dataset by identifying and collecting all the nouns from our dataset. This is
done by first preprocessing and tokenizing the extracted text from all the
screenshots present in our dataset and then performing a part-of-speech tagging
algorithm.  This tags each word with a word category such as noun, verb, and
adjectives.  Since we seek to identify the entities in our dataset, we collect
all the words tagged as noun and discard others, we consider the list of nouns
as entities. To filter out uncommon entities, we remove all entities that appear
in less than 10\% of the posts in our dataset. Next, we perform manual filtering
on this list to group together entities with the same meanings such as
\textit{vaccinations}, \textit{vaccines}, and \textit{shots}. This totals to 243
entities that have appeared in 10\% of the posts or more.

\parait{Entity sentiment analysis.} Next, we measure the sentiment towards these
entities in the user's posts. To this end, we utilize Google's Cloud Natural
Language API to perform entity sentiment analysis. The API yields a sentiment
score between -1 and +1 representing the overall emotion towards the particular
entity and a magnitude score representing the strength of the emotion. We set
the range for clearly positive sentiment at greater than 0.5 and clearly negative
sentiment at less than -0.5. We process all the extracted text from screenshots
with at least one of the identified entity present and calculate the sentiment
towards the entity. Finally, for our purposes, we compute the average sentiment
towards each entity from all of its mentions.

\para{Results.} We identify most common entities to include \textit{covid,
vaccine, virus, masks, news, government, shot}, and \textit{news} with their
mentions making up total of 32\% of all entity mentions. From our entity
sentiment analysis, we observe religious entities such as \textit{Jesus, amen,
lord,} and \textit{church} to have the highest percentage of positive sentiment
(82\% - 79\%). This can be explained by the presence of non-
followed by entities such as \textit{Ivermectin, truth, and
country}. The entities with the highest percentage of negative mentions were
\textit{mandates, businesses, china, government, masks, president, Biden,
vaccines}, and \textit{Fauci}. Distributing the entities across Covid-19
misinformation themes identified in \Cref{sec:narratives:themes}, we observe the
entities to match the themes such as government related entities being the most
common in anti-government posts with high values for negative sentiments.

\subsection{Takeaways}
In studying the content of the posts, we identify a significant dominance of
government related posts which suggests some level of politicization of COVID-19
in our dataset. Further inquiry in the progression of the narratives by users
suggest narratives to be centered more around reactive narratives to resist
recommendations, mandates and policies by the authorities as we observe in our
cohort analysis. The cohort analysis further confirms the dominance of political
topics in the dataset and the slight separation between conspiratorial topics
and mainstream political topics. 

\section{Sources of misinformation}
\label{sec:sources}
In this section, we focus on uncovering the role of public figures
(\Cref{sec:sources:public}) and external domains (\Cref{sec:sources:domains})
in the origination and amplification of misinformation themes.

\subsection{Public figures as sources of misinformation}
\label{sec:sources:public}
Public figures on social media platforms have a significantly different role in
an information cascade. Prior research highlights the importance of elites and
thought leaders to generate and propagate information in cascades. As explored
by Zhang \etal \cite{context-zhang}, users are more likely to consider verified
public figures to be more credible and trustworthy. Therefore, we seek to
answer the question: {\em How do public figures interact with misinformation
themes and their progression?}
At a high-level, we answer this question by extracting all posts made by
Facebook `verified users' that are shared by victims in our dataset. We then
obtain communities of verified users that frequently co-occur in our dataset.
We then analyze the themes of misinformation from verified users by the user's
occupation and political leaning (obtained from external sources).

\para{Methods.} We now outline our methods for extracting posts of verified
users shared by victims, obtaining occupation and political leanings of
verified users, and identifying communities of verified users.

\parait{Extracting posts of verified users shared by victims.}
To identify and measure the involvement of public figures in the narratives we
first need to identify whether a public figure is present in the screenshot of
the post. In this section, we outline the methods used to identify a public
figures presence in a post and collect their defining characteristics such as
their occupation and partisanship. First, to identify whether a public figure
is present in a screenshot of a post we use Facebook's verified badge label as
an identifier. Facebook, along with other platforms, places a verified badge
next to the names of verified public figures in their posts. The intent of the
verified badge is to ensure the post is authored by an account which is
verified to be actually controlled by the public figure. Since Facebook assigns
the verified badges only to public figures, we use the badges as a proxy for
identifying public figures. To identify whether a verified badge is present in
a screenshot, we use SIFT template matching algorithm. SIFT template matching
algorithm is a computer vision technique to identify whether a template, which
in our case is the verified badge, is present in an image. We perform the SIFT
template matching algorithm on all the screenshots in our collection and curate
a set of screenshots that contain at least one verified badge. Next, using
a combination of image analysis, to locate the text next to the verified badge,
and OCR, to extract the text, we extract the name of the public figures present
in the screenshot.

\parait{Obtaining occupation and political leaning of verified users.}
To assign attributes to each of the public figure, we search for their name on
Wikipedia using the Wikipedia API. For each public figure, we find their
occupation and partisanship from their Wikipedia page summary manually.  The
occupation of a public figure allows us to explore their credibility and
influence over their audience while their partisanship, if any, enables us to
identify partisan sources in Covid-19 misinformation. 

\parait{Identifying communities of verified users.}
Finally, to measure the involvement of public figures in the progression of
users narratives and beliefs, we construct an undirected graph of the posts
with public figures to measure a user's involvement with them and their
compatibility with each other. We connect the public figures, represented as
nodes, in the graph with edges representing the volume of users that have
shared both of the public figures posts. Using this graph, we extract
`communities' of public figures that were shared by a cohort of users. Similar
to our identification of theme progressions, this is
done using the Louvain community identification algorithm \cite{louvain}. 

\para{Results.}
In our collection of 20.8K unique posts, we identified 200 unique public
figures whose content was shared by our victims. In total, these accounted for
approximately 2.5\% of all our victims' posts. The most frequently shared 
figures included right-wing political commentators Tucker Carlson, Candace
Owens, Tomi Lahren, and Ben Shapiro who accounted for at least 20\% of all
shared posts.
Breaking down these figures by their derived occupations and political
leanings, we observe an obvious pattern (shown in \Cref{tab:figures}) ---
political elites (commentators and politicians) on the right-wing were most
likely to be the source of the posts shared by our victims.

\begin{table}[]
  \centering
  \resizebox{\columnwidth}{!}{%
    \begin{tabular}{@{}llllll@{}}
    \toprule
    \textbf{Category}              & \textbf{Most frequent figure}          & \textbf{Left}
    & \textbf{Neutral} & \textbf{Right} & \textbf{Total Unique} \\ \midrule
    Political Commentator & Tucker Carlson    & 1    & 1       & 195   & 42     \\
    Politician            & Ted Cruz          & 4    & 0       & 100   & 52     \\
    News Network          & Fox News          & 12   & 16      & 65    & 45     \\
    Advocacy Group        & Turning Point USA & 1    & 0       & 55    & 22     \\
    Celebrity             & Gina Carano       & 0    & 39      & 1     & 17     \\
    Health Agency         & CDC               & 0    & 9       & 0     & 6      \\ \bottomrule
    \end{tabular}
  }
  \caption{Number of posts including a public figure distributed by occupation
  and partisanship of the public figures in our dataset.}
  \label{tab:figures}
\end{table}

Next, studying the involvement of public figures in the identified
misinformation themes and progressions, we observe that posts with a public
figure were more likely to engage with anti-democrat and anti-government themes
compared to posts authored by the victims themselves (\ie non-shared posts).
The high frequency of these political themes from verified users can be
explained by the presence of high number of right-wing political elites. This
is shown in \Cref{tab:figures:topics}.
Looking simply at the types of posts made by verified users, we found that
posts including a public figure had significantly more focus on
anti-democratic and anti-immigration (\eg suggesting Covid-19 deaths were due
to an immigration crisis) themes compared to masks, conspiratorial narratives,
and anti-science claims. These results suggest that, at large, the political
elites were not very likely to peddle conspiracy or anti-science theories.
Instead, they focused on politicizing the pandemic and challenged perceptions
of trust in the government and it's institutions. 
Further, highlighting the power of elites and public figures, we observed that
victims who shared posts from public political figures showed a 22\%
(statistically significant) higher likelihood of also authoring similar posts
themselves (when compared to users who did not share posts from political
figures).

Our analysis using the undirected graph of the public figures show that it is
common for users to share mainstream public figures such as the overlap between
mainstream right-wing political commentators (Tucker Carlson, Ben Shapiro,
and Candace Owens) right-wing politicians, (Ted Cruz, Donald Trump Jr.) and
right-wing advocacy groups (PragerU, Turning Point USA) were significant enough
for the creation of a community. This shows that users who post mainstream
right-wing personalities are more likely to post or mention more mainstream
right-wing personalities. 
Other prominent cliques included groups of celebrities, satirical websites such
as BabylonBee, and advocacy groups. The formation of the right-wing mainstream
clique highlights an interesting fact that their misinformation themes are
similar and compatible enough (largely centered around anti-government and
political themes) to be shared by the same users. 
Finally, observing patterns around the progression of how content from verified
users is shared by victims, we observe two large cliques of victims. One clique
exclusively focused on sharing posts from different right-wing news networks
and the other exclusively focused on sharing posts from different right-wing
political commentators.

\begin{table}[]
  \centering
  \resizebox{\columnwidth}{!}{%
  \begin{tabular}{lrrrrrrrr}
   \toprule
  \multicolumn{1}{r}{\textbf{}} &
    \multicolumn{2}{c}{\text{\begin{tabular}[c]{@{}c@{}}Dist.\\ of posts\end{tabular}}} &
    \multicolumn{3}{c}{\text{\begin{tabular}[c]{@{}c@{}}\% Dist. of posts\\ by occupation\\ (Right)\end{tabular}}} &
    \multicolumn{3}{c}{\text{\begin{tabular}[c]{@{}c@{}}\% Dist. of \\ topics in
    each \\ group\end{tabular}}} \\
    \cmidrule(lr){2-3} \cmidrule(lr){4-6} \cmidrule(lr){7-9}  
  \textbf{Topics} &
    \textbf{Total} &
    \textbf{V} &
    \textbf{PC\%} &
    \textbf{P\%} &
    \textbf{NN\%} &
    \textbf{V\%} &
    \textbf{NV\%} &
    \textbf{\begin{tabular}[c]{@{}r@{}}Avg. \\ Effect\end{tabular}} \\ 
      \midrule
  \textit{Anti-government}  & 3040 & 181 & 27 & 29 & 14 & 23 & 58 & 0.35*  \\
  \textit{Anti-science}     & 750  & 9   & 22 & 22 & 0  & 6  & 3  & -0.03* \\
  \textit{Anti-vaccination} & 727  & 15  & 33 & 13 & 7  & 6  & 5  & -0.01  \\
  \textit{Masks}            & 564  & 3   & 33 & 33 & 0  & 5  & 1  & -0.04* \\
  \textit{Anti-Democrat}    & 544  & 53  & 74 & 4  & 4  & 4  & 17 & 0.13*  \\
  \textit{Religion}         & 532  & 4   & 25 & 25 & 25 & 4  & 0  & -0.03* \\
  \textit{Alt Solutions}    & 439  & 8   & 12 & 0  & 12 & 4  & 3  & -0.01  \\
  \textit{Mandates}         & 371  & 7   & 29 & 29 & 0  & 3  & 2  & -0.01  \\
  \textit{Immigration}      & 209  & 9   & 33 & 33 & 11 & 2  & 3  & 0.01   \\
  \textit{General-COVID}    & 170  & 2   & 0  & 0  & 0  & 1  & 1  & -0.01  \\
  \textit{Pro-Freedom}      & 134  & 0   & 0  & 0  & 0  & 0  & 0  & 0.00   \\
  \textit{Conspiracy}       & 105  & 1   & 0  & 0  & 0  & 1  & 0  & -0.01  \\
  \textit{Quarantine}       & 17   & 0   & 0  & 0  & 0  & 0  & 0  & 0.00   \\
  \midrule
  \textit{Asking for help}  & 883  & 2   & 0  & 0  & 0  & 7  & 1  & -0.07* \\
  \textit{Regret}           & 1174 & 5   & 20 & 40 & 0  & 10 & 2  & -0.08* \\
  \textit{Death}            & 2725 & 13  & 23 & 8  & 0  & 22 & 4  & -0.18* \\
  \hline
  \end{tabular}%
  }
  \caption{Distribution of topics in posts by verified users along with the \%
  distribution of users by each topic. The avg. effect column represents the
  increase in likelihood of a post being of that particular topic given a user
  is verified. {\bf V} and {\bf NV} denote verified and non-verified users,
  respectively. {\bf PC}, {\bf P}, and {\bf NN} denote political commentators,
  politicians, and news networks, respectively.
  * indicates statistically significant average treatment effect ($p<.05$).}
  \label{tab:figures:topics}
  \end{table}

\subsection{External domains as sources of misinformation}
\label{sec:sources:domains}

Similar to the public figures online, external domains have been studied to have a
significant influence on the reach of an information cascade. Tanaka \etal
\cite{tanaka2012transmission} show
how the inclusion of an external domain in a tweet makes it twice as more likely
to be retweeted suggesting that content with external domains are more likely to
be shared and propagate further in a cascade. Therefore, we seek to answer
the question: {\em How are external domains exploited for the propagation of
misinformation themes?} We answer this question by identifying all posts shared
by victims that include a link to an external domain. We then categorize these
domains by their reputations and analyze their involvement in victims'
progression through misinformation themes.

\para{Methods.} 
In order to analyze the involvement of external domains in a misinformation
theme, we first need to extract the external domains present in each screenshot
in our dataset and assign meaningful attributes to the domains. 
First, to identify whether a screenshot of a post contains a URL, we take
the text of the post and find any URL in the text using URL matching regular
expressions. Following the identification and extractions of our URL, we isolate
the domain part of the URL and discard the rest. 
Next, we identify the attributes of the domain which inform us about its
reputation and authenticity using Media Bias/Fact Check (MBFC)
\cite{huitsing2018newsweek}. We use MBFC to obtain the political bias and
factual reporting frequency for each domain and correlate these with
misinformation themes and progressions.
Finally, similar to our public figure analysis, we create a graph to explore
the compatibility of external domains.

\begin{figure}
    \centering
    \includegraphics[width=0.9\linewidth]{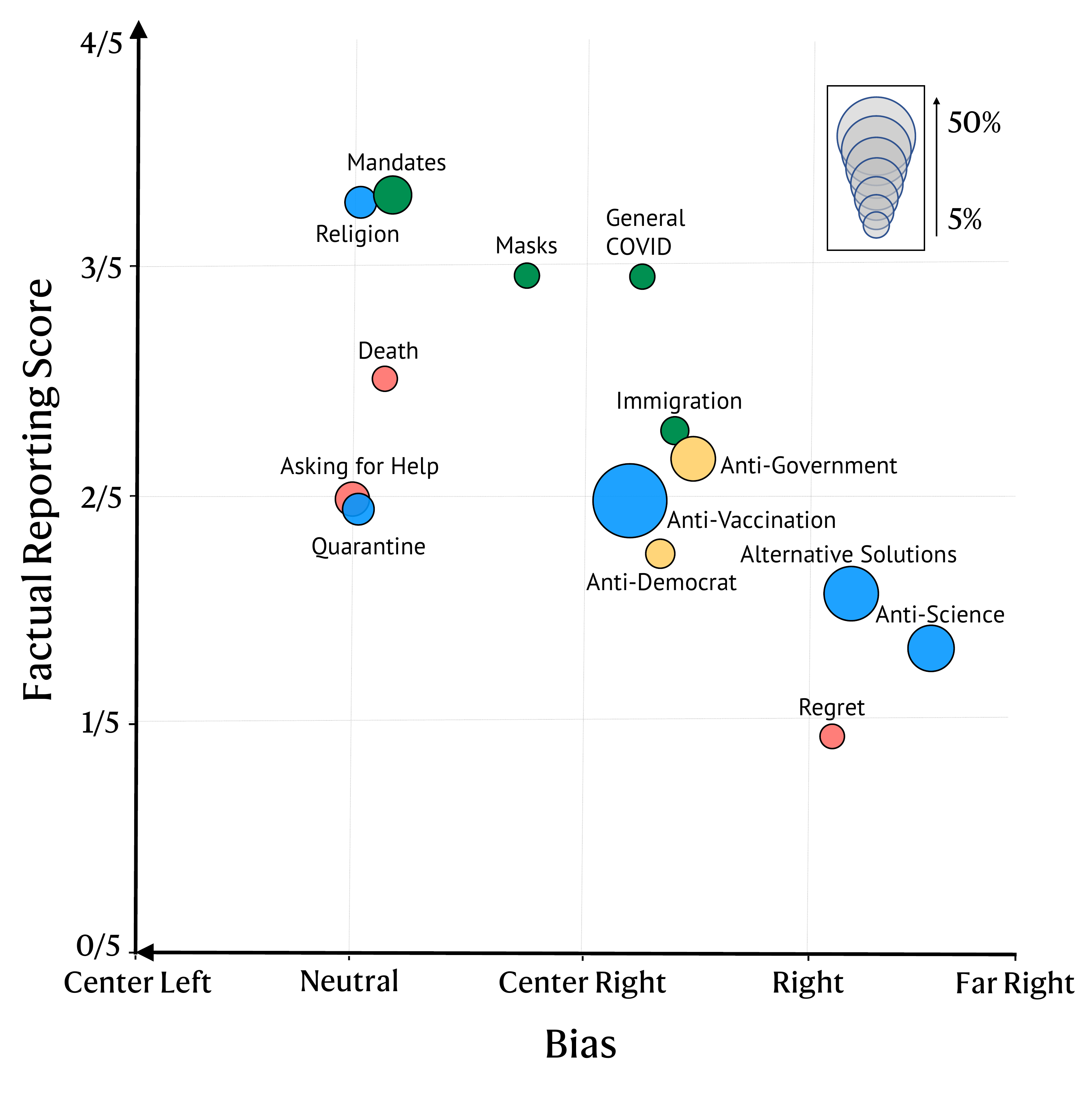}
    \caption{Percentage of posts with external domains in each topic category
    along with average bias and factual reporting score of the domains present.
    The size of the bubble is proportional to the percentage of posts with the
    topic containing an external domain.}
    \label{fig:H2.verified-narratives}
\end{figure}

\para{Results.}
In our dataset, we identify a total of 2.5K domains (925 unique domains).
Notably, 49\% of the victims in our dataset did not share a single link to an
external domain --- \ie all their misinfomation posts were from within
Facebook.
The most frequently observed domains were from video streaming platforms such as
Youtube, Rumble, and Bitchute. 
News-related domains such as Fox News and The Epoch Times accounted for a total
of 24\% of the URL-containing posts. These formed the basis of our MBFC
attribute analysis.
We observed that these shared external news domains contained a significant
amount of bias and misinformation. In our data, only 4\% of the domains had no
bias, while 20\% had center-left alignment, and 6\% had center-right alignment.
The remaining domains were right (40\% of all domains) or far-right aligned
(29\% of all domains). 
In other words, exploring the domains being shared in our dataset, we observe
a significant skew towards right and far-right sources. Furthermore, we observe
a significantly positive correlation between the domain's alignment with the
right and lack of factual reporting. 
Contrasting our previous results related to public figures, as shown in
\Cref{fig:H2.verified-narratives}, conspiratorial and pseudoscience topics such
as anti-vaccination, alternative solutions and anti-science topics were
significantly more prevalent. 
This result suggests that external domains (often unmoderated) are more likely
to be used as a source to spread non-factual (anti-vaccination, alternative
solutions and anti-science) topics. 
In our analysis on whether there are any patterns of users progression through
the domains, we observe a significant decrease in factual reporting and an
increase in severity of bias towards the right as we progress through a user's
posts timeline.  This suggests, that users might start with slightly factual and
neutral sources and later progress to consume and spread less factual and more
bias content. 

\subsection{Takeaways}
Taken together, the results from this section underscore the prominent
involvement of right-wing political commentators, politicians and news
networks in the politicization of Covid-19 by their almost exclusive involvement
in political themed misinformation. Our dataset highlights the high percentage of
users who had adopted and were sharing politicized narratives surrounding
Covid-19 that reflected their harmful beliefs. Our analysis on the external
domains also confirms the misinformation, especially related to anti-vaccination
beliefs, alternative solutions to Covid-19 and claims refuting the science of
Covid-19, was exceedingly being sourced from biased and non-credible external
domains which often contained blogs and videos from unregulated platforms.

\section{Moderation of Misinformation}\label{sec:platforms}

During the Covid-19 pandemic, in Feb. 2021, Facebook announced an expanded
effort to improve moderation of Covid-19 misinformation. Specifically targeted
themes were related to anti-vaccination, conspiracy theory, and anti-science
misinformation \cite{rosen2020update}.
These moderation efforts included outright removal of content or the
application of `soft moderation' warning labels on content.
In this section, we focus on evaluating the completeness
(\Cref{sec:platforms:completeness}) and consistency
(\Cref{sec:platforms:consistency}) of this effort.

\subsection{Completeness of interventions}\label{sec:platforms:completeness}
We now evaluate the fraction of misinformation posts in our dataset that
received a soft moderation intervention. Next, we analyse the application of
these interventions broken down by the misinformation theme to identify
specific gaps in moderation.

\para{Methods.}
We are unable to measure the posts deleted by Facebook, so we restrict
ourselves to measuring the application of soft moderation interventions on
misinformation posts shared by victims. The visible soft moderation
interventions are limited to flagging the content with one of the following
labels: 

\begin{enumerate}
    \item \textit{False Information.} This flag means the information found in
    the post is categorically false. Facebook state's content flagged as false
    information experiences dramatic reduction in its distribution and strong
    warning labels. 
    \item \textit{Partly False Information.} Content labeled with this flag
    include some factual inaccuracies. The interventions performed on content
    labeled partly false information are less severe compared to content labeled
    false information.
    \item \textit{Context Missing.} Posts labeled with this flag have the
    potential to mislead its readers without additional context. The
    interventions towards this type of content are minimal and are limited to a
    warning label stating that context is missing from the post.
    \item \textit{More Information Required.} Posts with this label are not
    identified as misleading or misinformation rather are provided with a
    redirect to information center to access more information regarding
    Covid-19. Posts labeled with this flag have mentions of vaccines and
    Covid-19 related topic.
\end{enumerate}

To identify whether a screenshot contains a post that is flagged and extract
which type of label has been flagged we perform template matching with text.
Using the extracted text from each of the post we search for string templates
created for each of the label above. Finally, we compute the distribution of
each type of label for each misinformation theme and source.

\begin{figure}[t]
    \includegraphics[width=\linewidth]{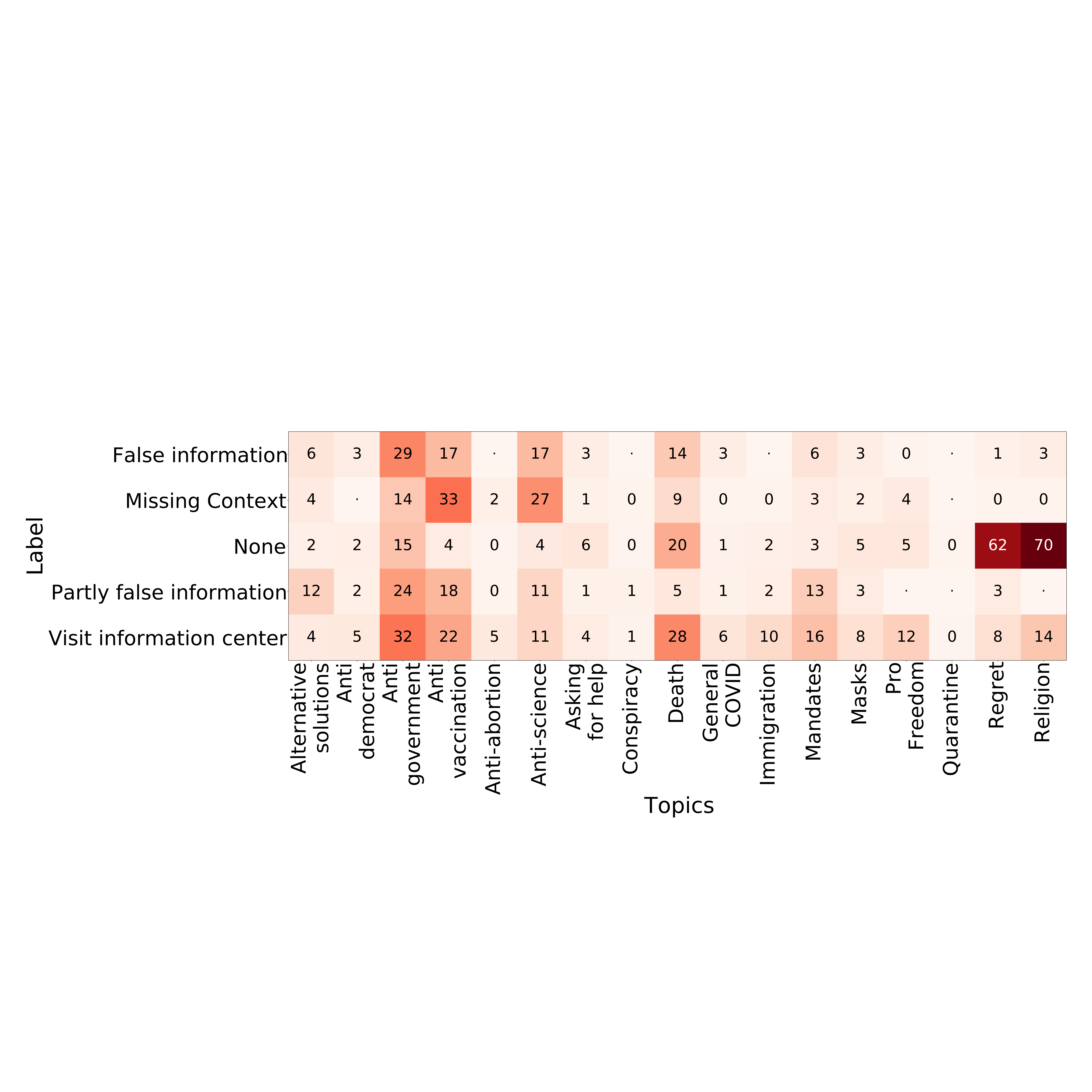}
    \caption{Percentage of misinformation labels present in each topic.}
    \label{fig:H3R1.misinformationclasses}
\end{figure}

\para{Results.}
In our dataset, we found only 6\% of the posts to be flagged by one of the
labels --- suggesting that Facebook's efforts at curbing Covid-19
misinformation was far from adequate or complete. 
Out of these labeled posts, 145 were labeled as false information, 133 were
labeled missing context, 114 were labeled partly false information and 889
were labeled to redirect the user to more information. 
In \Cref{fig:H3R1.misinformationclasses} we show the distribution of the labels
in each class.
Here we see the `False Information'  label being primarily distributed among
anti-vaccination, anti-government, and anti-science --- suggesting these themes
are more likely to be screened by Facebook fact checkers. Switching dimensions,
we see that the anti-government and anti-vaccination themes are also the most
frequent subjects of any type of soft moderation. Of concern, however, is that
alternative solutions and anti-science posts are often not the subject of any
interventions.

The most common entities present in posts labeled as `False Information' or
`Partly False Information' are \textit{vaccines}, \textit{media},
\textit{science}, \textit{masks}, \textit{CDC}, and \textit{virus} with these
entities accounting for 53\% of all the labels. Studying the distribution of
misinformation labels over verified users, we do not observe any significant
presence and find only 3 unique posts by right-aligned political commentators to
ever receive a `False Information' label. However, we do observe a significant
increase of 6\% likelihood of experiencing a label `redirecting to the
information center' for posts from verified users. Next, we observe containing
an external domain in a post significant increases the likelihood of being
labeled with a misinformation label. In total, we observe, out of the 2.5K URLs
mentioned, 38 containing `False Information' labels, 34 containing `Missing
Context' labels and 34 containing `Partly False Information' labels accounting
for 28\% of all misinformation labels. The distribution of misinformation on
external domains further increases as factual reporting decreases and the bias
skews towards far right with the fari-right URL's having the highest percentage
of misinformation labels (10\%) compared to other alignments.
%

\subsection{Consistency of interventions}\label{sec:platforms:consistency}
Consistent content moderation is central for effective moderation. In this
section, we use our limited dataset whether Facebook's interventions were 
consistent.

\para{Methods.}
To identify whether Facebook's soft moderation were consistent across posts with
the same content in our dataset, we use a combination of text analysis and
manual validation. By identifying all the posts with misinformation labels in
our dataset, we search for posts with similar keywords in our dataset. Following
this collection we validate, using the misinformation label string templates
and manual validation, the consistency of the application of Facebook
misinformation label.

\para{Results.}
We identify 42 posts labeled false or misleading while also having nearly
identical textual/semantic content to other posts in our dataset. We
find the labeling of misinformation to be consistently applied in all posts
sharing highly similar content. One key difference between the posts, however,
was the design of the misinformation label. While some posts had a superimposed
label partially obstructing the content of the post, others had no obstruction
and had the label present at the bottom of the post. This finding suggests the
presence of a moderation tool to identify and flag posts similar to previously
moderated posts.

\subsection{Takeaways}
Our analysis, albeit on a small dataset, suggests that Facebook's interventions
are consistent but far from complete. Specifically, these findings suggest that
Facebook's misinformation moderation efforts: (1) are most targeted towards
very specific types of anti-vaccination and anti-government themes --- largely
ignoring others, (2) do not appear to hold public figures to a higher standard
than regular users, (3) is adept at flagging content from problematic external
domains, and (4) utilize a mechanism to flag posts similar to already moderated
posts.

\section{Related Works}
Our work makes three key contributions operating on a unique dataset of posts made by
Facebook users with harmful beliefs related to Covid-19. We identify and characterize the
themes and sources these users engaged along with the reaction of Facebook in
the form of soft moderation actions. In this section, we place our work among
other studies performed on Covid-19 misinformation campaigns across different
social media platforms.

\para{Misinformation themes.}
Misinformation has been a long-standing problem in the social media space
and understanding the narratives and themes that drive misinformed beliefs has become a key area of
research. The Covid-19 narratives found on Twitter
\cite{jiang2020political,havey2020partisan} give support to our results from
\Cref{sec:narratives:themes} where we observe misinformed Facebook users to have
politicized the Covid-19 discussion. The authors, in studying the discussion
related to Covid-19 on Twitter, find political narratives to be the most
prominent theme and additionally for conservative users to be more engaged in conspiratorial and
political narratives. The authors argue the politicization of Covid-19
discussion to be a key reason for the dire consequences stemming from being distracting from actual health
and pandemic related discussion. Finally, in their work identifying the
information being propagated by bots on Twitter related to Covid-19 \cite{ferrara2020types}, the
author finds bots promoting already present pro-freedom alt-right ideologies
and spreading conspiratorial narratives. We find the narratives the bots, motivated by
an agenda to spread misinformation, spread and promoted most frequently on
Twitter matched with the Covid-19 themes identified and  propagated by
questionable external domains in our analysis of Facebook posts.

\para{Misinformation sources.}
Research in understanding the sources of information has been a key area of
understanding how misinformation propagates online. The role elite users and
external domains play in establishing trust, credibility and authority in
information cascades has been well studied across different platforms.
Specifically, related to misinformation on Covid-19, research done by Muric
\etal show the prominent role low credibility sources play in the spread of
Covid misinformation \cite{muric2021covid}. Studying a dataset of anti-vaccine
stance posts from Twitter, Muric \etal observe the most common links to be from
low credibility media sources and being shared by right-leaning accounts. This
observation is supported by our results from \Cref{fig:H2.verified-narratives}
where we observe anti-vaccination claims to be accompanied by external domains
of low credibility and having a right-leaning bias. A study on general early
Covid-19 narratives on Twitter by Singh \etal, however, show the significantly
overwhelming presence of high quality media sources compared to low quality
sources. Presence of high quality media sources in early general narratives of
Covid-19 compared to high presence of low quality sources found in our dataset
specifically in anti-vaccination narratives of Covid-19 confirms our observation
of the role of external domains in the spread of harmful beliefs on Covid-19 in
misinformed users. Exploring the role of verified users on Twitter, Andrews
\etal observe the power official accounts have in spreading and correcting
associated rumors \cite{andrews2016keeping}. Furthermore, Yang \etal show how
verified users on Facebook play a much significant role in spreading lower
credibility information compared to verified users on Twitter
\cite{yang2021covid}.  These results highlight the inaction of verified users
found in our dataset in correcting misinformation and spreading politically
polarized narratives.

\para{Platform interventions.}
Over the recent years, researchers have studied the effectiveness of content
moderation on social media platforms that include removal of content and
communities. Recently however, amidst the fears of fake news and misinformation,
social media platforms have started performing soft moderation interventions to
warn users about the accuracy of content. These soft moderation actions include
warning labels, overlays, and tags warning the user about the misrepresentation
or misinformation present in a post.
Research into understanding how effective these soft moderation interventions
have been in limiting the engagement and reach of misinformation highlights the
efforts still required to ensure consistency, completeness and effectiveness.
Mena et al., in their work \cite{mena2020cleaning} study the effectiveness
of Facebook's soft moderation interventions on misinformation. They find that
Facebook's soft moderation interventions are effective in reducing the likelihood
of a user sharing a post with a warning label. However, an empirical study by
Zannettou finds Twitter's warning labels to be not as effective especially for
Republican users \cite{zannettou2021won}. Additionally, they find Twitter's soft
moderation strategies to be inconsistent and often perceived as acts of
censorship. These results are further confirmed by Sharevski et al. in their
319 participant study to measure how users perceive Tweets with or without
warning labels \cite{sharevski2022misinformation}. Their analysis, focused on
COVID-19 misinformation, show people to resist warning labels especially if the
warning labels are not designed to cover the entire post. Additionally, their
study highlights the strongest predictor of perception of the accuracy of a post
to be the prior belief of a user rather than the presence of a warning labels.
They argue that extended use and misuse of warning labels might backfire in
misinformed users to completely ignore warning labels by perceiving the
moderator as being biased.
\section{Concluding Remarks} \label{sec:discussion}
\para{Limitations.}
Fundamentally, this work is a `best-effort' observational study aimed at better
understanding the role that the infodemic played in the large number of
Covid-19 deaths amongst the unvaccinated. Consequently, there are three
important limitations that influence our study and its findings.

\parait{Dataset limitations.}
Our reliance on crowd-sourced datasets, although necessary to overcome data
gathering limitations placed by social media platforms (Facebook, in
particular), introduce challenges to representativeness.
More specifically, we cannot ensure that the dataset contains a complete record
of all the misinformation shared by a victim. Since a victim's posts were
curated by other individuals for the purpose of cataloging in communities
engaging in schadenfreude, it is possible that not all misinformation-related
posts were recorded and a selection bias was introduced in the cataloging
process (\eg by selecting only the loudest anti-vaccination victims of Covid-19
for cataloging, or by selecting only specific types of posts from a victim).
Despite these challenges, we argue that our investigation and findings are
important because these curated datasets present a rare and unique opportunity
to understand the characteristics of misinformation encountered by the
misinformed victims of the Covid-19 pandemic. 
Further, the large number of unique victims and individual curators in our
datasets suggests that the general trends observed in our results must be
representative of a large segment of the US population. 

\parait{We can only uncover correlated relationships.} 
Because our study is observational, there is little opportunity to derive any
insight into causal relationships between misinformation themes (or entities)
and the victim's beliefs/passing.
This limits our conclusions to correlational relationships that suggest either
causation (victim adopted a belief because of a post) or rationalization
(victim made a post because it fit their beliefs). 
Additionally, our ability to run observational control-treatment analyses is
limited due to Facebook's scraping limitations that prevent us from creating
a meaningful control group.
Despite these limitations, our study is still able to draw conclusions
regarding the epistemes of misinformation (and their sources) that were most
likely to resonate with the eventual victims of Covid-19.

\parait{Text extraction may introduce errors.}
Finally, our analysis required text extraction from screenshots -- an already
noisy task further complicated by the cropping styles used by contributors,
the presence of mixed-media posts (\eg an image of text of varying fonts inside
a screenshot), and the importance of extracting sources of shared posts. 
At each step in our extraction pipeline, we conducted manual validation of
random samples to ensure high fidelity text extraction. Despite this effort,
unfortunately, we cannot guarantees of correctness over the entire dataset.

\para{Ethical considerations.}
Conducting this study was a challenging task, largely owing to the questionable
ethics of the dataset being studied and the communities that curated them.
In fact, there has been much media attention and criticism showered on these
communities for the unempathetic discourse surrounding the victims of
Covid-19 --- even spilling over to public conflicts between the moderators of
the communities \cite{Wapo-empathy-21}.
We undertook this work from the perspective that the victims cataloged by these
communities were ultimately failed by our political climate, leaders, and the
platforms they relied on. This paper is meant to highlight these failures so
they may not repeat. 
With regards to operational ethical considerations, our study did not scrape
posts from non-public domains and did not violate the scraping limitations set
by any platform/website. Whenever available, we relied on an official API for
data gathering and analysis. 

\para{Conclusions.}
This study suggests two key associations with the harms caused by the
infodemic: the politicization of Covid-19 by the political class and limited
moderation by online platforms. 
More specifically, our findings show that the anti-government theme of
misinformation propagated by the right-wing political and media ecosystem was
significantly more prevalent than anti-science and anti-vaccination themes,
amongst the victims of Covid-19 cataloged in our datasets. 
This result highlights the responsibility held by the political elites (and the
platforms that promote their voices) towards the masses and is complimentary to
the long line of work in the political sciences focused on showcasing the power
and authority of political elites in shaping public opinion in times of crisis
\cite{hutcheson2004us,jennings1992ideological,bachrach2017political}.
Our study also shows that Facebook largely failed to implement it's soft
moderation labels completely on Covid-19 misinformation, but maintained
consistency when it was applied. 
It remains unclear whether the limits of this application were by design (in
order to avoid causing `fatigue' and a reduction in label effectiveness
\cite{guo2020identity}) or due to challenges with scale
\cite{habib2022proactive}.
Regardless of the reason, it is paramount that platforms (or their regulators)
recognize their sociopolitical influence and redesign their
platforms/algorithms to more effectively dampen the reach of harmful (and in
this case, deadly) misinformation.

\balance

\bibliographystyle{aaai21}
\bibliography{memes}

\end{document}